\documentclass[twocolumn,showpacs,prl,superscriptaddress,floatfix]{revtex4}
\usepackage{graphicx}
\usepackage{dcolumn}
\usepackage{bm}
\usepackage{amssymb}
\usepackage{amsmath}
\usepackage{amsfonts}
\usepackage{amssymb}
\usepackage{hyperref}
\usepackage{color}
\usepackage{comment}

\begin{document}

\title{Holographic Storage of Biphoton Entanglement}
\author{Han-Ning Dai*}
\affiliation{Hefei National Laboratory for Physical Sciences at Microscale and
Departmentof Modern Physics, University of Science and Technology of
China,Hefei,Anhui 230026, China}
\author{Han Zhang*}
\affiliation{Hefei National Laboratory for Physical Sciences at Microscale and
Departmentof Modern Physics, University of Science and Technology of
China,Hefei,Anhui 230026, China}
\author{Sheng-Jun Yang}
\affiliation{Hefei National Laboratory for Physical Sciences at Microscale and
Departmentof Modern Physics, University of Science and Technology of
China,Hefei,Anhui 230026, China}
\author{Tian-Ming Zhao}
\affiliation{Hefei National Laboratory for Physical Sciences at Microscale and
Departmentof Modern Physics, University of Science and Technology of
China,Hefei,Anhui 230026, China}
\author{Jun Rui}
\affiliation{Hefei National Laboratory for Physical Sciences at Microscale and
Departmentof Modern Physics, University of Science and Technology of
China,Hefei,Anhui 230026, China}
\author{You-Jin Deng}
\affiliation{Hefei National Laboratory for Physical Sciences at Microscale and
Departmentof Modern Physics, University of Science and Technology of
China,Hefei,Anhui 230026, China}
\author{Li Li}
\affiliation{Hefei National Laboratory for Physical Sciences at Microscale and
Departmentof Modern Physics, University of Science and Technology of
China,Hefei,Anhui 230026, China}
\author{Nai-Le Liu}
\affiliation{Hefei National Laboratory for Physical Sciences at Microscale and
Departmentof Modern Physics, University of Science and Technology of
China,Hefei,Anhui 230026, China}
\author{Shuai Chen}
\affiliation{Hefei National Laboratory for Physical Sciences at Microscale and
Departmentof Modern Physics, University of Science and Technology of
China,Hefei,Anhui 230026, China}
\author{Xiao-Hui Bao}
\affiliation{Physikalisches Institut, Reprecht-Karls-Universitat Heidelberg,
Philosophenweg 12, 69120 Heidelberg, Germany}
\affiliation{Hefei National Laboratory for Physical Sciences at Microscale and
Departmentof Modern Physics, University of Science and Technology of
China,Hefei,Anhui 230026, China}
\author{Xian-Min Jin}
\affiliation{Hefei National Laboratory for Physical Sciences at Microscale and
Departmentof Modern Physics, University of Science and Technology of
China,Hefei,Anhui 230026, China}
\author{Bo Zhao}
\affiliation{Hefei National Laboratory for Physical Sciences at Microscale and
Departmentof Modern Physics, University of Science and Technology of
China,Hefei,Anhui 230026, China}
\affiliation{Institute for Theoretical Physics, University of Innsbruck, A-6020
Innsbruck, Austria}
\author{Jian-Wei Pan}
\affiliation{Hefei National Laboratory for Physical Sciences at Microscale and
Departmentof Modern Physics, University of Science and Technology of
China,Hefei,Anhui 230026, China}
\date{\today}

\begin{abstract}
Coherent and reversible storage of multi-photon entanglement with a
multimode quantum memory is essential for scalable all-optical quantum information processing.
Although  single photon has been successfully stored
in different quantum systems, storage of multi-photon entanglement remains challenging because of the
critical requirement for coherent control of photonic entanglement source, multimode quantum memory, and quantum interface
between them. Here we demonstrate  a coherent and
reversible storage of biphoton Bell-type entanglement with a holographic
multimode atomic-ensemble-based quantum memory. The retrieved biphoton entanglement violates
Bell's inequality for 1 $\mu$s storage time and a memory-process fidelity of
98\% is demonstrated by quantum state tomography.
\end{abstract}

\maketitle

\renewcommand{\thefootnote}{\fnsymbol{footnote}} \footnotetext[1]{%
These authors contributed equally to this work.} \renewcommand{\thefootnote}{%
\arabic{footnote}}

Faithfully mapping multi-photon entanglement into and out of quantum memory is of crucial importance for
scalable linear-optical quantum computation~\cite{knill01} and long-distance quantum communication~\cite{duan01}. Recently, storage of nonclassical light~\cite{Juls04,appel08} and single photons~\cite{chan05,eisa05,choi08,choi10,zhang11,clau11,sagl11,spec11} has been demonstrated in various quantum systems, such as atomic ensemble, solid system, and single atom.
Among these, the atomic-ensemble-based quantum memory holds the promise to
implement multimode quantum memory for multi-photon entanglement.
A natural method for this purpose is to select spatially separated sub-ensembles of a large
atomic ensemble as different quantum registers~\cite{lan09,choi10}, for which
the number of stored modes is limited by the spatial dimension of the atomic ensemble.
More powerful methods, such as exploring large optical depth of an atomic ensemble~\cite{nunn08,Zeu11}, or utilizing photon echoes~\cite{hoss09} or atomic frequency combs techniques~\cite{ried08}, have been employed to demonstrate atomic-ensemble-based multi-mode memories.

\begin{figure}[tbp]
\includegraphics[width=8cm]{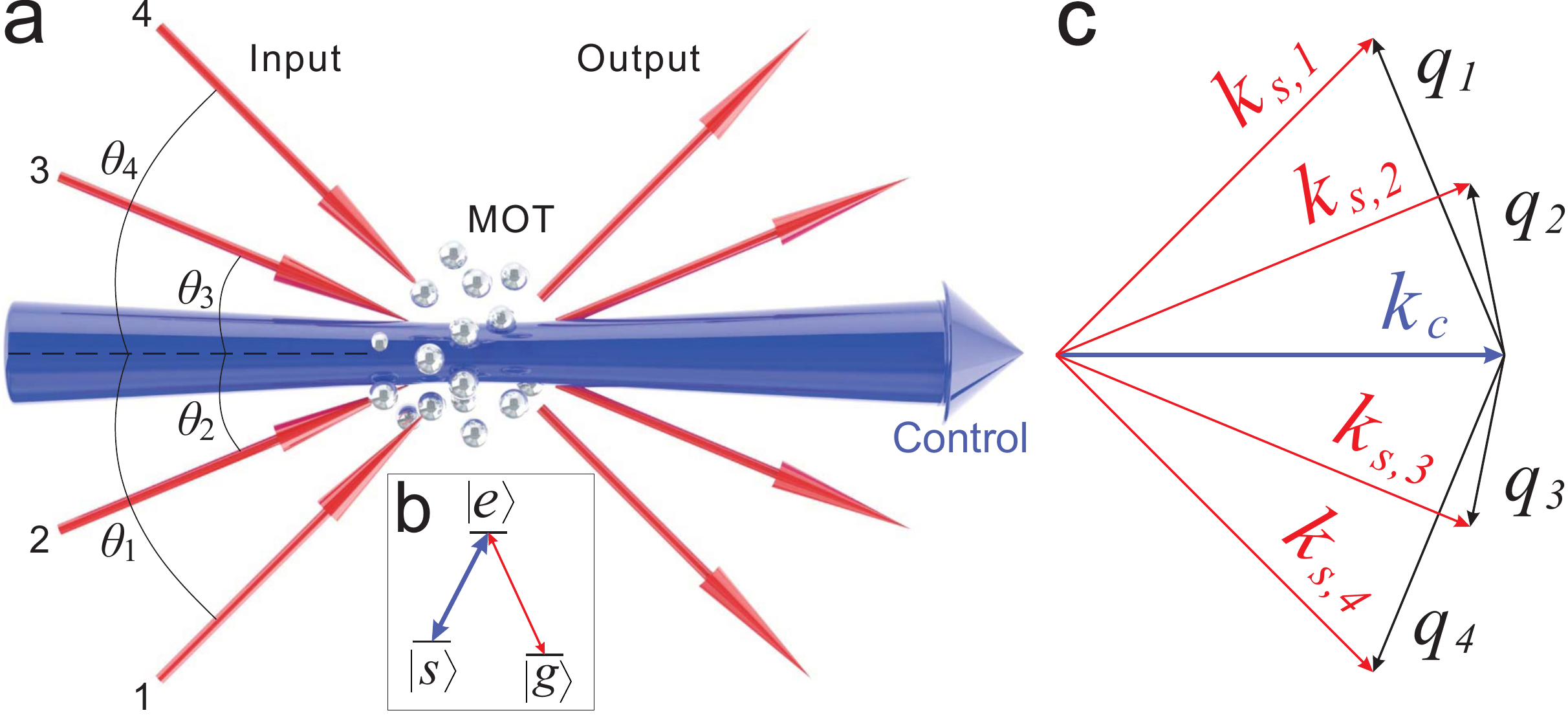}
\caption{\textbf{(a)} Schematic view of the quad-mode holographic quantum
memory. The control field is shined into the atomic cloud horizontally, and the signal
modes are incident from different directions in the same plane, with angles $(\protect\theta %
_{1},\protect\theta _{2},\protect\theta _{3},\protect\theta _{4})=(-1^{\circ
},-0.6^{\circ},0.6^{\circ },1^{\circ })$ relative to the the control field.
\textbf{(b)} The typical $\Lambda$-type energy levels, with two ground states $|g\rangle$ and $|s\rangle$ and an excited state $|e\rangle$.  The $|e\rangle- |g\rangle$ and the $|e\rangle-|s\rangle$
transition are coupled to the signal and control fields, respectively.  \textbf{(c)} Illustration of the wave vectors of
the spin waves. The input signal field with wave vector of $\mathbf{k}_{s,i} $ ($i=1,2,3,4$) is mapped to a spin wave with
wave vector $\mathbf{q}_{i}=\mathbf{k}_{s,i}-\mathbf{k}_{c}$, with $\mathbf{k}_{c}$ the wave vector of the
control field. }
\label{fig1}
\end{figure}

An alternative and elegant method is to implement the atomic ensemble as a
holographic multimode quantum memory~\cite{tord08,vas08,Surmacz08}, using spatially overlapped but orthogonal
spin waves as different quantum registers. For clarity, we illustrate this in the example of storing a single-photon
state in an atomic ensemble of $N$ atoms
that have two long-lived ground states $|g\rangle $ and $|s\rangle $~\cite{flei00,liu01,phil01,phi08}. Initially, a
``vacuum'' state $|\mathrm{vac}\rangle=|g_{1}...g_{N}\rangle$ is prepared such that all
the atoms are at the $|g\rangle$ state.
The single-photon state is then mapped into the ensemble as a collective state
$|1,\mathbf{q}\rangle =S_{\mathbf{q} }^{\dagger }|$vac$\rangle =(1/\sqrt{N})\sum_{j}e^{i\mathbf{q}\cdot \mathbf{x}%
_{j}}|g_{1}\cdots s_{j}\cdots g_{N}\rangle $, where $\mathbf{x}_{j}$ is the
position of the  $j$th atom, $S_{\mathbf{q}}^{\dagger }=(1/\sqrt{N}%
)\sum_{j}e^{i\mathbf{q}\cdot \mathbf{x}_{j}}|s\rangle _{j}\langle g|$ is the
collective creation operator of a spin wave with wave vector $\mathbf{q}$. For $N \gg 1$, one has $[S_{\mathbf{q}_{1}},S_{%
\mathbf{q}_{2}}^{\dagger }]\approx \delta _{\mathbf{q}_{1}\mathbf{q}_{2}}$, namely, the collective states satisfy the
orthogonality relation $\langle 1,\mathbf{q}_{1}|1,\mathbf{q}_{2}\rangle \approx \delta _{%
\mathbf{q}_{1}\mathbf{q}_{2}}.$ Therefore, one can encode different qubits by different phase patterns
and employ a single atomic ensemble as a holographic multimode quantum memory. Since the information is stored globally throughout the medium, one can achieve high-capacity data storage.
Recently, holographic storage of classical light and microwave pulses have been demonstrated~\cite{vud08,shu08,wu10}.

\begin{figure*}[t]
\includegraphics[width=16cm]{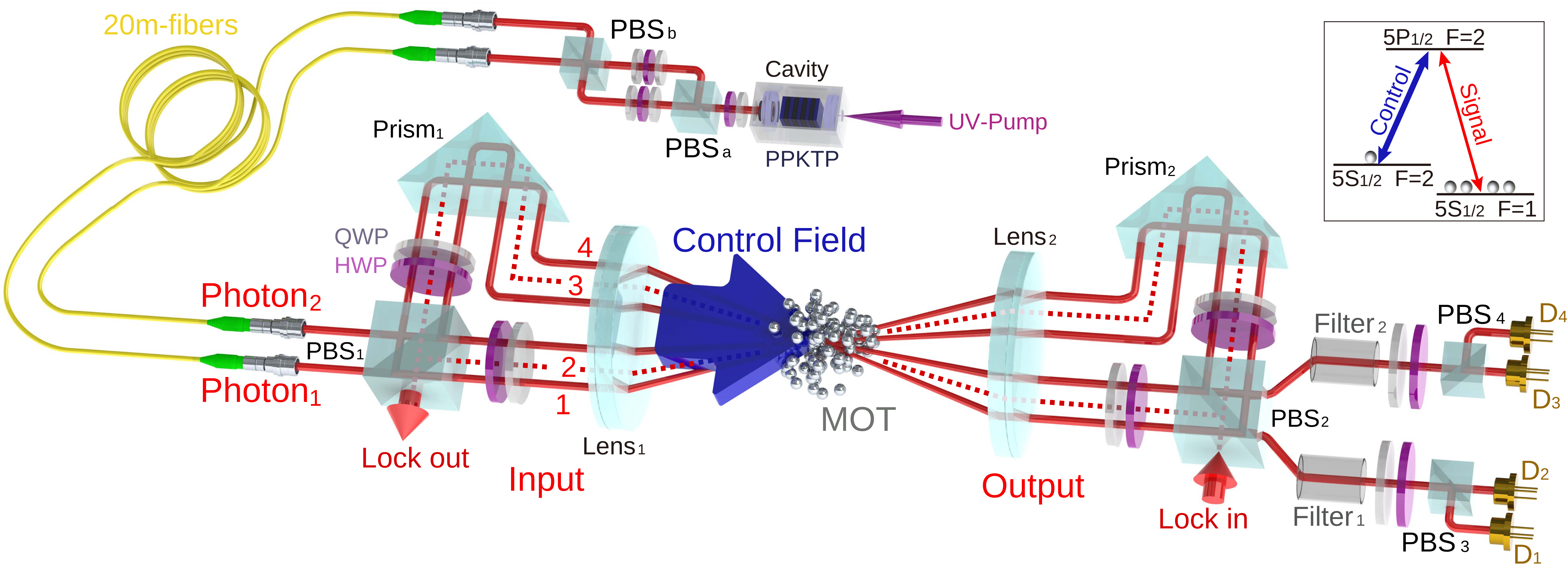}
\caption{Illustration of the experimental setup. The two
polarization-entangled photons produced by a narrowband SPDC source are
directed to the memory lab through 20-m fibers. Different polarization
components  are separated by the polarization beam splitter (PBS)
1, and are coupled to the quad-mode holographic quantum memory. All the
light fields are turned into right-hand circular polarization ($\protect\sigma%
^{+}$) by wave plates. The biphoton entanglement are stored into the four quantum registers by adiabatically
switching off the control light. After a controllable delay, the photons are
retrieved out, transferred back to their original polarization states by wave plates and combined on PBS$_{2}$. Then the retrieved entangled photons are guided into filters, containing a Fabry-Perot cavity and a hot atomic cell to
filter out the leakages from the control light, and detected by single photon
detectors. The path length difference between PBS$_{1}$ and PBS$_{2}$ is actively
stabilized by an additional phase beam (dashed line). Inset: The experimental laser lights and atomic levels.}
\label{fig2}
\end{figure*}

Here we report an experimental demonstration of holographic storage of
biphoton Bell-type entanglement with a single atomic ensemble, in which four orthogonal spin waves with different wave vectors are used as a quad-mode quantum memory.
The posterior biphoton entanglement is mapped
into and out of the quad-register holographic quantum memory,
via a technique based on electromagnetically induced transparency (EIT).
Violation of  Bell's inequality is observed for storage time up to 1 $\mu $s
and a memory-process fidelity of 98\%, calculated by quantum state tomography, is achieved.

The experimental scheme and setup are shown in Fig.~\ref{fig1} and~\ref{fig2}, respectively.
In the memory lab, we prepare, within 14 ms, a cold $^{87}$Rb atomic ensemble consisting of about $10^{8}$ atoms
in a dark Magnetic-Optical-Trap (MOT).
The temperature of the atomic cloud is about 140 $\mu $K, and the optical depth (OD) is about 10.
The typical $\Lambda$-type energy-level configuration is shown in Fig.~\ref{fig1}b,
where $|g \rangle$, $|e\rangle$, and $|s\rangle $ correspond
to the $^{87}$Rb hyper-fine states $|5S_{1/2},F=1\rangle$, $|5P_{1/2},F=2\rangle$, and $|5S_{1/2},F=2\rangle $, respectively. All the atoms are initially prepared at $|g\rangle $.

A strong classical control field couples $|e\rangle $-$|s\rangle $ transition with
wave vector $\mathbf{k}_{c}$ and beam waist diameter $w_{c}\approx 850$ $\mu $m,
while the to-be-stored quantum field, which has four components (see below), couples $|g\rangle $-$|e\rangle $ transition with beam waist diameter $w_{s}\approx 450$ $\mu $m.
The control field is focused at the ensemble center,
and the four components of the signal field are guided through the atomic cloud
along four different directions, which are in the same plane but
with different angles  $\theta _{i} \, (i=1-4)$  relative to the control-light direction~\cite{chens07}.
We set $ (\theta _{1},\theta _{2},\theta _{3},\theta _{4})=(-1^{\circ },-0.6^{\circ},0.6^{\circ },1^{\circ })$,
as illustrated in Fig. \ref{fig1}a. By carefully adjusting the directions of the control and signal beams, we make all the light modes overlap in the center of the atomic ensemble.  The atomic ensemble has a length $L\approx 2$ mm, and the
signal fields propagate within the control field during storage.

Each component $i$ of the signal field is associated with a wave vector $\mathbf{k}_{s,i}$,
and is to be stored in a spin wave with $\mathbf{q}_{i}=\mathbf{k}_{s,i}-\mathbf{k}_{c}$.
By careful alignment, a holographic quad-mode quantum memory, with approximately equivalent optical depth and similar performance, is established. We measure the EIT transmission spectrum and perform slow-light experiment. For a control light with a
Rabi frequency of about 7 MHz, we observe an EIT window of $2.2$ MHz and a
delay time of about $160$ ns for all the four modes. Note that such a holographic quantum memory is different from the scheme in Ref.~\cite{choi10,lan09}, where each signal mode requests a spatially separated atomic sub-ensemble.

The biphoton entanglement comes from a narrowband
cavity-enhanced spontaneous parametric down conversion (SPDC) entanglement source as in previous work~\cite{bao08,zhang11}.
The source cavity contains three main parts, i.e., a nonlinear crystal, a tuning crystal and an output coupler.
The nonlinear crystal is a 25-mm type-II a periodically
poled KTiOPO$_{4}$ (PPKTP) crystal, whose
operational wavelength $\lambda\approx795$ nm is designed to match the D1 transition line of $^{87}$%
Rb. The cavity is locked intermittently to a Ti: Sapphire laser using the Pound-Drever-Hall method. The linewidth and finesse of the cavity are measured to be 5 MHz and 170, respectively.

Polarization-perpendicular photon pairs are created by applying a ultraviolet (UV) pumping
light, which is up converted from the Ti: Sapphire laser. Single-mode
output is achieved by using a filter cavity (made of a
single piece of fused silica of about 6.35 mm) with a finesse of 30, which
removes the background modes. Polarization-entangled photon pairs are post-selected by
interfering the twin photons at polarization beam splitters (PBSs).  The ideal outcome state corresponds to a
Bell state
\begin{equation*}
|\phi ^{+}\rangle _{p}=(|H\rangle _{1}|H\rangle _{2}+|V\rangle _{1}|V\rangle
_{2})/\sqrt{2}
\end{equation*}%
with H(V) represents the horizontal(vertical) polarization of the photons. Under a continuous wave (CW) pump with a pump power of 4 mW, the spectrum brightness of the polarization-entangled pairs after the filter cavity is about 50 s$^{-1}$mW$^{-1}$MHz$^{-1}$. In the storage experiment, the entangled signal photons are created by a 200 ns pump pulse, which is cut from a 28 mW CW pump laser.  The production rate is about 33 s$^{-1}$. The measured ratio of counts under $|HH\rangle/|HV\rangle$ and $|++\rangle/|+-\rangle$ bases are 14.3:1 and 23.1:1, respectively, with $|\pm\rangle=(|H\rangle\pm|V\rangle)/\sqrt{2}$.

\begin{figure}[t]
\includegraphics[width=8cm]{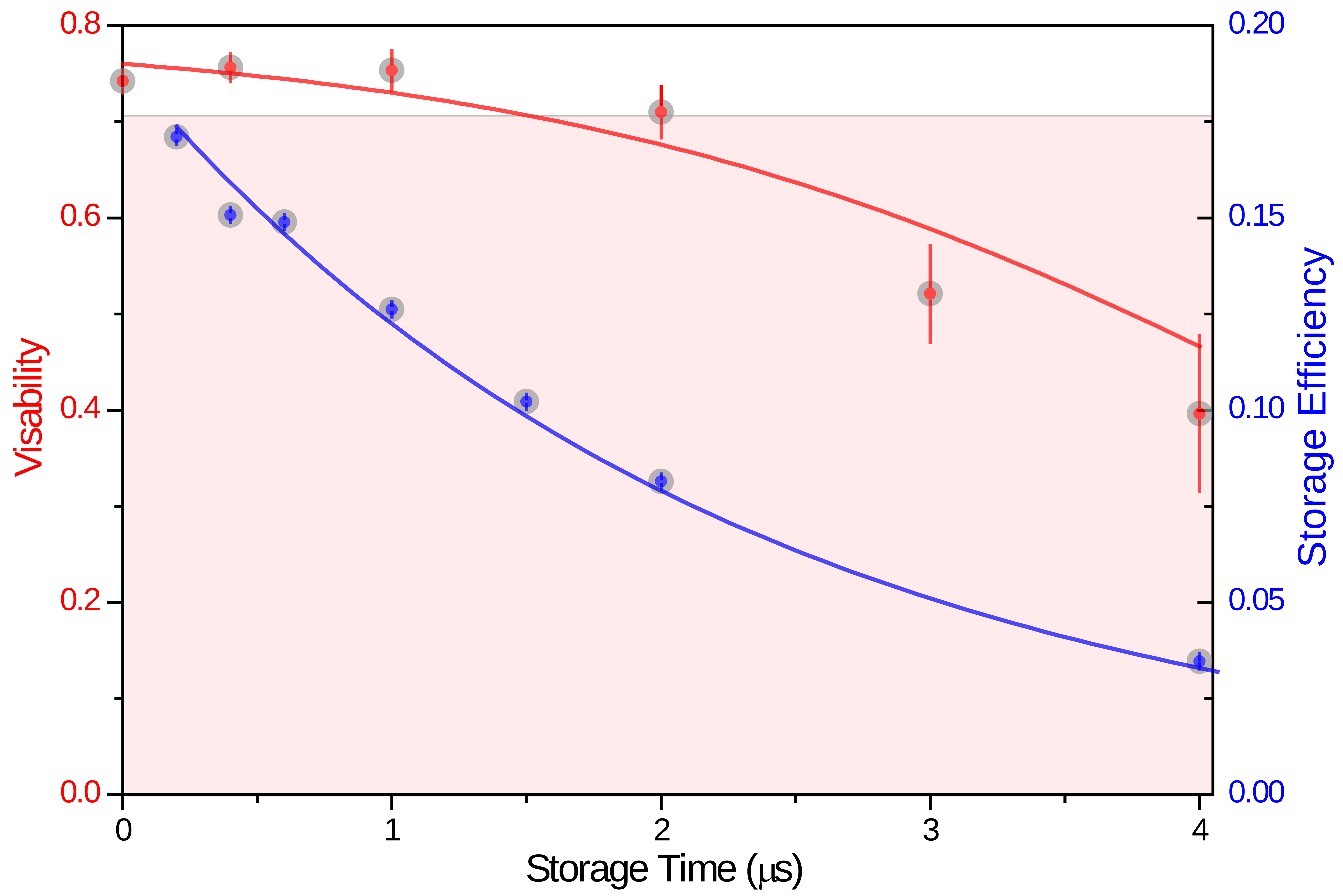}
\caption{ Average visibility (red, left axis) and overall storage efficiency (blue, right axis) of the retrieved biphoton state versus storage time. An exponential fitting (blue solid line) of the storage efficiency yields a lifetime of $\protect\tau=2.8\pm0.2$ $\protect\mu$s. The retrieved biphoton state is measured under $|H/V\rangle $, $|\pm\rangle $, and $|R/L\rangle $ bases with $|\pm\rangle=(|H\rangle
\pm |V\rangle)/\sqrt{2}$, and $|R/L\rangle=(|H\rangle\pm i|V\rangle)/\sqrt{2}$.
The average visibility is fitted using $V=1/(a+b e^{2t/\tau})$ (red solid line) with $a$ and $b$ the fitting parameters. The result shows that within about 1.6 $\protect\mu$s the visibility is above the threshold of 0.71 to violate
the CHSH-Bell's inequality. Error bars represent $\pm$ standard deviation}
\label{fig3}
\end{figure}

The signal photon pair is directed to the memory lab
with 20-meter single-mode fibers.
The different polarization components are  spatially separated by PBS$_{1}$, then transferred to right-hand circular polarized ($\sigma ^{+}$) by wave plates,
and then guided to the four quantum registers by lens (see Fig. \ref{fig2}).
More precisely, the  $|H\rangle _{1}$, $|H\rangle _{2}$,
$|V\rangle _{2}$, and $|V\rangle_{1}$ polarization components are coupled to mode 1-4, respectively.
After these components entering the atomic
ensemble, we adiabatically switch off the control light, and the photonic
entanglement is mapped into the atomic ensemble. This yields an
entanglement among the four quantum registers
\begin{equation*}
|\psi \rangle _{a}=(S_{\mathbf{q}_{1}}^{\dagger }S_{\mathbf{q}_{2}}^{\dagger
}+S_{\mathbf{q}_{3}}^{\dagger }S_{\mathbf{q}_{4}}^{\dagger })|\text{vac}%
\rangle /\sqrt{2}
\end{equation*}%

After a controllable delay, we adiabatically switch on the control light and
convert the atomic entanglement back into photonic entanglement. The
polarization  states of the output photons are transferred back linearly
polarized by wave plates and combined by PBS$_{2}$ to reconstruct the biphoton entanglement.
The two retrieved entangled photons are respectively guided into a
filter consisting of a Fabry-Perot cavity (transmission window 600 MHz) and
a pure $^{87}$Rb vapor cell with atoms prepared in $|5S_{1/2},F=2\rangle $,
and then detected by single-photon detectors. The measured overall average storage
efficiency is shown in Fig. \ref{fig3}, which yields a
$1/e$ lifetime of $2.8\pm0.2$ $\mu $s. The measured coincidence rate without storage and after 1 $\mu$s storage time is 1.3 s$^{-1}$ and 0.03 s$^{-1}$, respectively.
The propagating phase between PBS$_{1}$ and PBS$_{2}$ is actively stabilized within $\lambda_{l} /30$ by an
additional phase lock beam with $\lambda_{l}\approx780$ nm~\cite{chens07,zhang11}.

To verify that the biphoton Bell-type entanglement is faithfully mapped into
and out of the four holographic quantum registers, we first measure the
retrieved biphoton state in $|H/V\rangle $, $|\pm \rangle$, and $|R/L\rangle =(|H\rangle \pm i|V\rangle )/\sqrt{2} $ bases at different storage time.
The average visibility is shown in Fig. \ref{fig3}, which for storage time less than 1.6 $\mu $s,
exceeds the threshold 0.71 to violate CHSH-Bell's inequality.
Note that the reduction of the visibility with storage time is mainly due to the background coincidences caused by the dark counts and the leakage from the control field. We further measure the correlation function $E(\phi _{1},\phi _{2})$, with $\phi
_{1}(\phi _{2})$ the polarization angle for signal photon 1(2), and
calculate quantity $S=|-E(\phi _{1},\phi _{2})+E(\phi _{1},\phi_{2}^{\prime })+E(\phi _{1}^{\prime },\phi _{2})+E(\phi _{1}^{\prime},\phi_{2}^{\prime })|,$ where $(\phi _{1},\phi_{1}^{\prime },\phi _{2},\phi_{2}^{\prime })=(0^{\circ },45^{\circ },22.5^{\circ},67.5^{\circ }).$ We obtain $S=2.54\pm 0.03$ for the input state, and $S=2.25\pm 0.08$ for the retrieved state after 1 $\mu $s storage. The violation of the CHSH-Bell's inequality ($S>2$~)\cite{clauser69}
confirms the entanglement has been coherently and reversibly stored in the
quad-mode holographic quantum memory.

\begin{figure}[t]
\includegraphics[width=8cm]{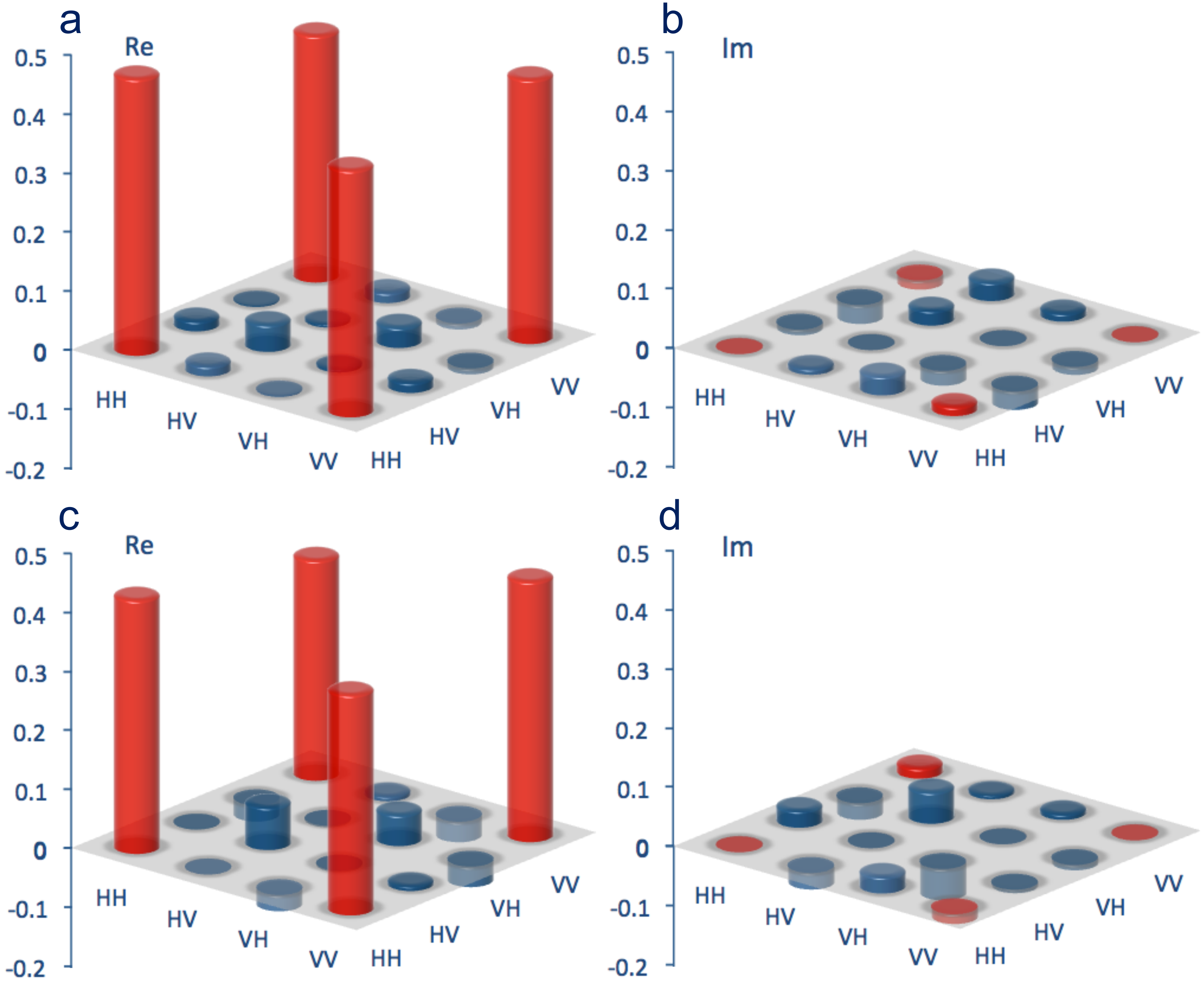}
\caption{ Density matrix of the input state (\textbf{a}, \textbf{b}) and of the output state after 1 $\mu$s storage (\textbf{c}, \textbf{d}), obtained from quantum state tomography. \textbf{a} and \textbf{c} are for the real parts, and \textbf{b} and
\textbf{d} are for the imaginary parts.}
\label{fig4}
\end{figure}

To quantitatively assess the fidelity of the storage process, we perform the
quantum state tomography~\cite{Ariano03,kwiat05} to construct the density matrix $\rho_{\rm{in}}$ of the input
and $\rho_{\rm{out}}$ of the output state after 1 $\mu$s storage, in which the polarization state of each photon is measured with two single-photon
detectors under different detection settings.
The results are illustrated in Fig. \ref{fig4}, from which the fidelity of the measured state $\rho$
on the ideal Bell state $\rho _{\phi ^{+}}$ is calculated as $F(\rho_{\phi^+}, \rho)
=(\mathrm{Tr}\left(\sqrt{\sqrt{\rho_{\phi^+} }\rho \sqrt{\rho_{\phi^+}}}\right))^{2} $. A Monte Carlo simulation technique~\cite{kwiat05} is applied to calculate the uncertainties of the fidelity. Briefly, an ensemble of 100 random sets
of data are generated according to Possionian distribution and then the density matrices are obtained by means of the maximum likelihood method. This yields a distribution of fidelities, from which the mean value and uncertainties of the fidelity are calculated. We obtain $F(\rho_{\phi^{+}},\rho _{\mathrm{in}})=(87.9\pm0.5)\%$ for the input state $\rho_{\mathrm{in}}$
and $F(\rho_{\phi^{+}},\rho_{\mathrm{out}})=(81\pm2)\%$,
beyond the threshold~\cite{zeil03} 78\% for Werner states to violate Bell's inequality. The fidelity of the memory process is given by $F(\rho _{\text{in}},\rho _{\text{out}}) =(98.2\pm0.9)\%$.

In summary, we have experimentally demonstrated  the coherent mapping of
a biphoton Bell-type entanglement, created from a narrowband SPDC source, into and out
of a four-register holographic quantum memory, with a high memory-process
fidelity of 98\% for 1 $\mu $s storage time. The narrowband photonic entanglement source inherits the advantage of conventional broadband SPDC source, and can be used to generate multi-photon entanglement beyond biphoton entanglement. A novel feature of the holographic quantum memory
is that one can use more modes by simply choosing the directions of the signal and control fields.
The memory capacity $N_m$ for a coplanar configuration may be estimated by the geometric mean of the Fresnel numbers of the illuminated regions as $N_m\sim w_{c}w_{s}/(\lambda L)$~\cite{Surmacz08}, which is $N_m\sim240$ for our experimental parameters. Increasing the beam waist diameters or extending to a three-dimensional geometry would allow much more modes.
Individual control of each quantum register may be achieved by using an optical cavity and
employing the stimulated Raman adiabatic passage technique~\cite{tord08} or employing the phase match method~\cite{Surmacz08}.

To extend our work to storage of multi-photon entanglement, we have to improve the brightness of the entanglement source, and increase the retrieval efficiency and lifetime of the quantum memory.
The storage efficiency is about 15\%, which can be improved by further
increasing the optical depth and reducing the linewidth of the narrowband
entanglement source. The storage time is about 1 $\mu $s, which is limited by inhomogeneous broadening induced by residual magnetic field, and can be improved to  be of order of millisecond
by trapping the atoms in optical lattice and using the magnetic-insensitive
state~\cite{zhaob09,zhaor09}. Our work opens up the possibility of
scalable preparation and high-capacity storage of multi-photon entanglement,
and  also sheds light on the emerging field of holographic quantum information processing.

This work was supported by the National Natural Science Foundation of China,
the National Fundamental Research Program of China (grant no. 2011CB921300),
the Chinese Academy of Sciences, the Austrian Science Fund, the European
Commission through the European Research Council Grant and the Specific
Targeted Research Projects of Hybrid Information Processing.

\end{document}